\documentclass[twocolumn,aps,prl,10pt,amsmath,amssymb,nofootinbib,showpacs,superscriptaddress,floatfix]{revtex4-1}

\DeclareFontFamily{U}{rcjhbltx}{}
\DeclareFontShape{U}{rcjhbltx}{m}{n}{<->rcjhbltx}{}
\DeclareSymbolFont{hebrewletters}{U}{rcjhbltx}{m}{n}

\usepackage{graphicx}
\usepackage{color}
\usepackage[usenames,dvipsnames]{xcolor}
\usepackage[colorlinks=true,linkcolor=Red,citecolor=Green,linktoc=page]{hyperref}
\usepackage{multirow}
\usepackage{float}
\usepackage{flushend}
\usepackage{balance}
\usepackage[varg]{txfonts}
\usepackage{ulem}
\usepackage{fancyhdr}

\DeclareMathSymbol{\lamed}{\mathord}{hebrewletters}{108}

\begin{document}
\title{Oblique confinement states are superinsulators, not topological insulators}

\begin{abstract}

\end{abstract}



\author{C.\,A.\,Trugenberger}

\affiliation{SwissScientific Technologies SA, rue du Rhone 59, CH-1204 Geneva, Switzerland}


\
\begin{abstract}
We show that the oblique confinement states in the Cardy-Rabinovici model are superinsulators, not topological insulators. This is because their only bulk excitations are strings that completely prevent the separation of $\pm$ charges, causing a strictly infinite resistance (below a critical temperature and/or voltage). This is very different from the typical activated behaviour of topological insulators, caused by a finite bulk energy gap. 

\end{abstract}
\maketitle


Superinsulators \cite{dst, vinokurnature, dtv1} (for a review see \cite{book}) are materials whose electrodynamic response is given by Polyakov's compact quantum electrodynamics (QED) action in the monopole condensation phase \cite{polyakov, polyakovbook}. As such they are described by a string theory, which goes by the name of confining string theory \cite{quevedo1, polyakovconf}. The excitations consist of a massive photon and strings. When probe charges are inserted in the material, the strings induce a linear potential between them, which causes a diverging resistance, i.e. confinement, in samples larger than the string scale for all temperatures below a finite transition temperature and voltages below a critical value. This behaviour arises naturally in inhomogeneous superconductors and has been experimentally detected in planar films of various superconductors with emergent granularity (for a review see \cite{book}), in which case the monopoles are instantons rather than solitons \cite{polyakovbook} and form a plasma rather than a condensate. 

In 3D, the compact QED action can be supplemented by the topological $\theta$-term of axion electrodynamics \cite{axion}, leading to the Cardy-Rabinovici model \cite{cardy}, in which the confinement phases are replaced by {\it oblique confinement}, caused by the condensation of dyons carrying both magnetic and electric charge via the Witten effect \cite{witteneffect}. The corresponding states of matter where introduced in \cite{pseudo} under the natural name of {\it oblique superinsulators}, to reflect their confining nature. In particular, it was first shown there that there is a topological limit in which only edge excitations survive. This limit was recently re-discovered and prompted the labelling of these states as oblique topological insulators \cite{fradkin}. It is the purpose of this brief note to review the physics of oblique superinsulators and to show that they are not topological insulators. 

The Cardy-Rabinovici model \cite{cardy} is given by the Euclidean partition function 
\begin{eqnarray}
Z &&= \sum_{ \{ n_{\mu} , \ S_{\mu \nu} \} }\int {\cal D} A_{\mu} {\rm e}^{-S} \ ,
\nonumber \\
S &&= \int d^3 x {1\over 4e^2} \left( F_{\mu \nu} -2\pi S_{\mu \nu} \right)^2 + i A_{\mu} \left( n_{\mu} + {\theta \over 2\pi} m_{\mu}  \right)  \ ,
\label{cr}
\end{eqnarray}
where $n_{\mu}$ and $m_{\mu}= (1/2) \epsilon_{\mu \nu \alpha \beta} S_{\alpha \beta} $ are singularities representing electric and magnetic currents, respectively and $e$ is the usual dimensionless coupling constant. The second dimensionless parameter $\theta$ is an angle and the minimal coupling of the magnetic monopoles, endowing them with charge $e\theta /2\pi$, is called the Witten effect \cite{witteneffect}. The electric and magnetic currents are to be considered as dynamical degrees of freedom, to be summed over in the partition function. To do so rigorously, of course requires formulating the model on a lattice. The first term in (\ref{cr}) is Polyakov's famed compact QED model, in which electric charges are confined for sufficiently strong coupling \cite{polyakovbook}. 

There are three types of possible phases in the Cardy-Rabinovici model, a superconducting phase, when the electric charge condenses, a Coulomb phase, when no charge condenses and, finally, various {\it oblique confinement} phases when excitations carrying both electric and magnetic charge, so called dyons, condense. Importantly, if the (electric, magnetic) charge combination $(n, m)$ condenses, all excitations with quantum numbers not in this condensate are confined. In particular, if $m\ne 0$,  electric charge is always confined. This is why these oblique confinement phases were originally identifed as {\it oblique superinsulators} \cite{pseudo}. The same phases were recently posited as oblique topological insulators \cite{fradkin}. However, confinement is clearly not the behaviour of an insulator, topological or not. An insulator has vanishing bulk resistance only at zero temperature, at finite temperatures it has an activated bulk behaviour governed by an energy gap.  As we now explain in more detail confinement means that the material has infinite bulk resistance even at finite temperatures, up to a critical temperature and/or voltage above which current starts to pass. This behaviour is the exact dual to a superconductor, for which the resistance is strictly zero below a critical temperature and/or current. 

The Cardy-Rabinovici model can be obtained \cite{fate} from an effective field theory involving a one-form gauge field $a_{\mu}$ and a two-form gauge field $b_{\mu \nu}$ coupled by the topological BF term (for a review see \cite{blau}). However, contrary to the 2D case, in which all kinetic terms for the fields in the topological action are power-counting irrelevant and do not influence the infrared (IR) properties of the theory, in 3D the kinetic Maxwell term for the one-form field is marginal and must be added to the effective field theory. The three-form kinetic term for $b_{\mu \nu}$  (for a review see \cite{blau}) remains instead irrelevant and must be added only as a gauge-invariant ultraviolet (UV) regulator with mass $\Lambda$. Of course, the kinetic Maxwell term spoils the topological character of the effective action, which is reflected in the fact that the Cardy-Rabinovici model is not topological. Contrary to 2D, one cannot conclude a priori that the IR limit of the theory will be a topological model, one must derive where the dimensionless coupling $e$ flows when the cutoff is removed, $\Lambda \to \infty$. 

The effective action for the Cardy-Rabinovici model in the phase with a monopole condensate (carrying automatically also electric charge due to the Witten effect) can be easily written down using the Julia-Toulouse prescription, as adapted to generic antisymmetric tensor gauge fields in \cite{quevedo1}. The Julia-Toulouse prescription involves promoting the original electromagnetic field tensor $F_{\mu \nu} = \partial_{\mu} A_{\nu} -\partial_{\nu} A_{\mu}$ to a fundamental two-form $B_{\mu \nu}$, the additional degree of freedom describing the fluctuations of the condensate. Correspondingly, the original current $j_{\mu}$ coupling to $A_{\mu}$ has to be promoted to a surface coupling to $B_{\mu \nu}$, 
\begin{eqnarray}
Z &&= \int _{{\rm surfaces}} \int
{\cal D}B_{\mu \nu } \exp \left( -S +i \int _{\rm surface}
B_{\mu \nu }\ d\sigma _{\mu \nu } \right) \ ,
\nonumber \\
S &&= \int d^4x \  {1\over 12 \Lambda^2} H_{\mu \nu \alpha}H_{\mu \nu \alpha} 
+ {1\over 4e^2} B_{\mu \nu }B_{\mu \nu }  + i{\theta \over 32 \pi^2} B_{\mu \nu} \epsilon_{\mu \nu \alpha \beta} B_{\alpha \beta} \ ,
\label{confstring}
\end{eqnarray}
where $H_{\mu \nu \alpha} = \partial_{\mu} B_{\nu \alpha} + \partial_{\nu} B_{\alpha \mu} + \partial_{\alpha} B_{\mu \nu}$ and the UV regulator mass scale $\Lambda \propto \sqrt{z} $ with $z$ the monopole fugacity. This is the action of the confining string \cite{quevedo1, polyakovconf, quevedo}. When no external charges are present the sum is over closed surfaces, when external probe charges are inserted the sum is over all open surfaces whose fixed boundaries describe the original currents $j_{\mu}$ of the probe charge-hole pair. Charges and holes are thus bound by a string, the dual of an Abrikosov vortex in a superconductor. 

Note that, as first derived in \cite{pseudo}, in the formal limit $\Lambda \to \infty$ and $e \to \infty$, this model becomes topological since only the term $B_{\mu \nu} \epsilon_{\mu \nu \alpha \beta } B_{\alpha \beta}$ survives in the action and, correspondingly, there are edge modes on a manifold with boundaries. This is what lead the authors of \cite{fradkin} to conclude that the oblique confinement states are actually oblique topological insulators. However, because of the non-topological marginal term present in the effective action in 3D, the rigorous IR limit must be defined by following the flow of the dimensionless coupling $e$ when the UV cutoff is removed, $\Lambda \to \infty$. As we now show, when this is done, the model is neither topological nor an insulator. 

To do so we first note that, in presence of the $\theta$ parameter, the photon mass gap, describing the inverse length scale on which Coulomb interactions are screened becomes \cite{quevedo1, quevedo, pseudo}
\begin{eqnarray}
m_{\theta} &&= {e\Lambda \over 4\pi} \sqrt{\left( {4\pi \over e^2} \right) + t^2} \ ,
\nonumber \\
t &&= {\theta \over \pi} \ .
\label{mass}
\end{eqnarray}
The string tension, however, is given by \cite{quevedo, book} 
\begin{equation} 
\tau = {\Lambda ^2 \over 4\pi} {{\rm e}^{-{et\over 4\pi}} \over \sqrt{et\over 4\pi}} \ .
\label{ten}
\end{equation} 
Because of the exponential factor, for strong coupling we have
\begin{equation}
\sqrt{\tau} \ll m_{\theta} \ .
\label{rel}
\end{equation}
Since the inverse square root of the string tension describes the typical string length, while the photon mass is its typical width, this 
means that, when $e$ becomes large, the strings are much longer than wide. Actually, the model can be renormalized by removing the cutoff $\Lambda \to \infty$, in which case $e$ also flows to infinity, $e\to \infty$ with the string tension $\tau$ fixed \cite{quevedo}. Of course, in this limit the photon mass $m_{\theta} \to \infty$ and the resulting model describes fundamental strings (no thickness) with a spin dependent on $\theta$, realizing the original Polyakov conjecture that a spin of the string, encoded in the self-intersection number of the Euclidean world surface would have exactly this stabilizing effect \cite{polyakovbook}. This is the phenomenon of dimensional transmutation: because of non-perturbative effects, the original dimensionless coupling $e$ is replaced by a dimensionful parameter, here the string tension $\tau$. 

The resulting IR model has two types of excitations, edge excitations and bulk strings of typical energy scale $\sqrt{\tau}$. This is not a topological model, the string action crucially depends on the metric. Nor is it an insulator. The insulating bulk gap $m_{\theta}$ has been renormalized to infinity and the physical bulk string excitations prevent any charge transport below a critical temperature or a critical applied voltage which reduce the effective string tension to zero. In this regime the resistance is strictly infinite and the state is thus a {\it superinsulator}, a dual superconductor. Of course, if one chooses to consider exclusively energies below the string scale, only the topological edge excitations survive \cite{pseudo}. Nothing is changed, however to the fact that the bulk still has strictly infinite resistance and is thus not an insulator but a superinsulator.


\begin{thebibliography}{10}
	\expandafter\ifx\csname url\endcsname\relax
	\def\url#1{\texttt{#1}}\fi
	\expandafter\ifx\csname urlprefix\endcsname\relax\def\urlprefix{URL }\fi
	\providecommand{\bibinfo}[2]{#2}
	\providecommand{\eprint}[2][]{\url{#2}}
	
\bibitem{dst}
M. C. Diamantini, P. Sodano, C. A. Trugenberger, Gauge theories of Josephson junction arrays. 
\textit{Nuclear Physics}\,B\,\textbf{474}, 641 -- 677 (1996).

\bibitem{vinokurnature}V. M. Vinokur \textit{et~al.} Superinsulator and quantum synchronization.
\textit{Nature} \textbf{452}, 613 -- 615 (2008). 

\bibitem{dtv1}
M. C. Diamantini, C. A. Trugenberger, V. M. Vinokur, Confinement and asymptotic freedom with Cooper pairs. \textit{Comm. Phys.}\,\textbf{1}, 77 (2018). 	
	
\bibitem{book} C. A. Trugenberger, Superinsulators, Bose metals and high-$T_c$ superconductors: the quantum physics of emergent magnetic monopoles. World Scientific, Singapore (2022). 

\bibitem{polyakov}A. M. Polyakov, Compact gauge fields and the infrared catastrophe. {\it Phys. Lett.} {\bf 59}, 82-84 (1975).  

\bibitem{polyakovbook} A.\,M.\,Polyakov, {\it Gauge Fields and Strings}, Harwood Academic Publisher, Chur (Switzerland) (1987).

\bibitem{axion}F. Wilczek, Two applications of axion electrodynamics. {\it Phys. Rev. Lett.} {\bf 58}, 1799 (1987). 

\bibitem{cardy} J. L. Cardy and E. Rabinovici, Phase structure of $Z_p$ models in the presence of a $\theta$ parameter. {\it Nucl. Phys. } {\bf B205} 1-16 (1982).

\bibitem{witteneffect} E. Witten, Dyons of Charge $\theta /2\pi$. {\it Phys. Lett.} {\bf 86}, 283-287 (1979). 

\bibitem{pseudo}M. C. Diamantini, C. A. Trugenberger, V. M. Vinokur, Topological Nature of High Temperature Superconductivity, {\it Advanced Quantum Technologies} 2000135 (2021).

\bibitem{fradkin}B. Moy, H. Goldman, R. Sohal and E. Fradkin, Theory of oblique topological insulators, arXiv:2206.07725v1. 

\bibitem{fate} M. C. Diamantini and C. A. Trugenberger, Topological superconductivity, topological confinement and the vortex quantum Hall effect, {\it Phys. Rev.} {\bf B84} 094520 (2011). 

\bibitem{blau} D. Birmingham, M. Blau, M. Rakowski, and G. Thompson, Topological field theory, {\it Phys. Rep.} {\bf 209}, 129 
(1991).

\bibitem{quevedo1} F. Quevedo and C. A. Trugenberger, Phases of antisymmetric tensor field theories. {\it Nucl. Phys.} {\bf B501} 143-172 (1997).


\bibitem{polyakovconf}A. Polyakov, Confining strings. {\it Nucl. Phys.} {\bf B486} 23-33 (1997). 


\bibitem{quevedo}  M. C. Diamantini, F. Quevedo and C. A. Trugenberger, Confining Strings with Topological Term. {\it Phys. Lett.} {\bf B396} 115-121 (1997).

\end{thebibliography}
\end{document}